\documentclass[sigconf]{acmart}
\usepackage{booktabs,multirow}
\usepackage{balance} 

\copyrightyear{2020}
\acmYear{2020}
\setcopyright{rightsretained}
\acmConference[AIES '20]{Proceedings of the 2020 AAAI/ACM Conference on AI, Ethics, and Society}{February 7--8, 2020}{New York, NY, USA}
\acmBooktitle{Proceedings of the 2020 AAAI/ACM Conference on AI, Ethics, and Society (AIES '20), February 7--8, 2020, New York, NY, USA}
\acmDOI{10.1145/3375627.3375820}
\acmISBN{978-1-4503-7110-0/20/02}

\usepackage{graphicx}

\begin{document}
\fancyhead{}

\title{Saving Face: Investigating the Ethical Concerns of Facial Recognition Auditing}

\author{Inioluwa Deborah Raji}
\affiliation{
  \institution{University of Toronto}
  \country{Canada}}
\email{deborah.raji@mail.utoronto.ca}

\author{Timnit Gebru}
\affiliation{
  \institution{Google Research}
  \country{United States}}
\email{tgebru@google.com}

\author{Margaret Mitchell}
\affiliation{
  \institution{Google Research}
  \country{United States}}
\email{mmitchellai@google.com}

\author{Joy Buolamwini}
\affiliation{
  \institution{MIT Media Lab.}
  \country{United States}}
\email{joy.buolamwini@gmail.com}

\author{Joonseok Lee}
\affiliation{
  \institution{Google Research}
  \country{United States}}
\email{joonseok@google.com}

\author{Emily Denton}
\affiliation{
  \institution{Google Research}
  \country{United States}}
\email{dentone@google.com}

\begin{abstract}
Although essential to revealing biased performance, well intentioned attempts at algorithmic auditing can have effects that may harm the very populations these measures are meant to protect. This concern is even more salient while auditing biometric systems such as facial recognition, where the data is sensitive and the technology is often used in ethically questionable manners. We demonstrate a set of five \textit{ethical concerns} in the particular case of auditing commercial facial processing technology, highlighting additional design considerations and ethical tensions the auditor needs to be aware of so as not exacerbate or complement the harms propagated by the audited system. We go further to provide tangible illustrations of these concerns, and conclude by reflecting on what these concerns mean for the role of the algorithmic audit and the fundamental product limitations they reveal.
\end{abstract}

\maketitle

\section{Introduction}
\label{sec:intro}

Facial processing technology (FPT) is a broad term that encompasses a variety of tasks ranging from face detection, which involves locating a face within a bounding box in an image; facial analysis, which determines an individual's facial characteristics including physical or demographic traits; and face verification or identification, which is the task of differentiating a single face from others. 

FPT can be deployed for a wide range of uses ranging from smiling detection to gauge customer satisfaction, to estimating the demographic characteristics of a subject population, to tracking individuals using face identification tools~\cite{pnp}. Corporate rhetoric on the positive uses of this technology includes claims of ``understanding users'', ``monitoring or detecting human activity'', ``indexing and searching digital image libraries'', and verifying and identifying subjects ``in security scenarios''~\cite{Face19,MSFT19,Cla19,Amz19}. 

The reality of FPT deployments, however, reveals that many of these uses are quite vulnerable to abuse, especially when used for surveillance and coupled with predatory data collection practices that, intentionally or unintentionally, discriminate against marginalized groups~\cite{benjamin2019race}. The stakes are particularly high when we consider companies like Amazon and HireVue, who are selling their services to police departments or using the technology to help inform hiring decisions respectively \cite{Ref48,Ref54}. 

Civil rights organizations have already sounded the alarm against facial recognition technology in particular and the need for urgent policy and regulatory action to restrict its use. Several states in the United States--California, Washington, Idaho, Texas, and Illinois--in addition to some cities--San Francisco, Oakland, and Somerville--are already taking the lead in regulating or outright banning the use of these technologies through coordinated campaigns such as the ACLU's Community Control Over Police Surveillance (CCOPS) initiative. As of the writing of this paper, federal bill proposals such as the Algorithmic Accountability Act~\cite{AAA19}, Commercial Facial Recognition Privacy Act of 2019~\cite{CFRT19} and No Biometric Barriers Act~\cite{BB19} have also been proposed in the U.S., as well as a bill proposing a moratorium in the U.K. \cite{UKbill}. 

Several of these proposals and their corresponding memos explicitly recommend that the results of FPT audits such as Gender Shades \cite{Joy1} and benchmarks developed through the National Institute of Standards and Technology \cite{Ref34} serve as conditions for FPT accreditation or moratorium. The language used in these proposals frames such audits as trusted mechanisms to certify the technology as safe and reliable for deployment. 

In this paper, we caution against this stance, outlining ethical concerns we have identified in the development and use of these algorithmic audits. We believe these concerns to be inherent restrictions to the utility of these audits within the broader evaluation of these systems, and propose to explicitly acknowledge these limitations as we make use of the audits in practice and in policy. 

Our primary contributions are as follows. We first develop CelebSET, a new intersectional FPT benchmark dataset consisting of celebrity images, and evaluate a suite of commercially available FPT APIs using this benchmark. 
We then use our benchmark and audit development process as a case study,
and outline a set of ethical considerations and ethical tensions relevant to algorithmic auditing practices. 

\section{CelebSET: A Typical FPT Benchmark}

Datasets such as Face Recognition Vendor Tests (FRVT), IARPA Janus Benchmarks (IJB)~\cite{Ref34}, and the Pilot Parliaments Benchmark (PPB)~\cite{Joy1} have been key to identifying classification bias in the specific tasks of face detection, verification, identification, identity clustering and gender recognition. Audits conducted using these datasets have heavily informed many of the recent attempts to address the dearth of diversity in FPT benchmarks and proposed frameworks for evaluating classification bias in deployed systems \cite{2019arXiv190110436M,srinivas2019face,Ref48}. Although these prior studies evaluate commercially deployed models for gender and racial discrimination in FPT, there are few benchmarks that enable the evaluation of the full range of classification tasks commercially available in facial processing technology today, and none that enables an intersectional black box audit evaluation for such a wide range of tasks. 

To address this gap, we develop a benchmark called CelebSET, a subset of the IMDB-WIKI dataset \cite{imdbwiki} that includes 80 celebrity identities--20 of the most photographed celebrities from each of the subgroups `darker male'(DM), `darker female'(DF), `lighter male'(LM) and `lighter female'(LF). Metadata pertaining to celebrity ethnicity were crawled from the celebrity fan websites FamousFix.com and ethiccelebs.com, and indexed to celebrity identities from the IMDB-WIKI dataset.`Darker'(D) is approximated by taking a subset of the images of celebrities tagged with the crawled `Black' label from our crawled dataset, and selecting subjects within the 3 to 6 range of the Fitzpatrick skin type scale \cite{sachdeva2009fitzpatrick} using the reported Fitzpatrick labels of celebrity references for guidance. The `Lighter'(L) individuals are a subset of subjects with the  crawled `White' label, also verified through visual inspection to approximate the skin type of celebrity references for a Fitzpatrick scale of 1 to 3. Gender - defined as `Male'(M) and `Female'(F) - as well as age metadata is taken from the IMDB-WIKI dataset. We estimate the age of the celebrity depicted in each photo by subtracting the person's known date of birth from the timestamp of the uploaded photo. We manually identify 10 smiling and 10 non-smiling example images for each celebrity. 

We use the original uncropped images with bounding box labels provided from the IMDB-WIKI dataset~\cite{imdbwiki} to perform our audit on the detection task and use cropped face images from the original dataset to audit facial analysis tasks. 
The full dataset with meta-data, cropped and uncropped images is available as Supplementary Materials. 




\subsection{API Evaluation on CelebSET}

We evaluate the APIs of Microsoft, Amazon, and Clarifai, which offer the widest scope of facial analysis tasks. Microsoft is notable as a target corporation in the initial Gender Shades study \cite{Joy1}, and Amazon in the follow up study, Actionable Auditing \cite{AA}. Clarifai is notable as an API that has not been previously included in any prior audit studies. For tasks such as automatic gender recognition, smile detection, and name identification, we evaluate the accuracy of the predicted value as compared to ground truth. For age prediction, we allow for an 8-year acceptance margin to accommodate the age range results of the Amazon API. We evaluate detection performance using the $AP_{50}$, the average precision at a threshold of 0.50 intersection over union (IoU). All evaluation results are as of October 2019. Calculation details, code and complete results of the audit are included in Supplementary Materials.

\subsubsection{Overall Performance}
As shown in Table~\ref{tab:over-acc}, all APIs perform best on the task of gender classification--with the exception of Clarifai which performs best on face detection. It is noteworthy that two of the target APIs, Amazon and Microsoft, have been publicly audited for gender classification in previous studies, and have since released new API versions in response to the audits citing improvements \cite{Joy1,AA}. All APIs perform worst on age classification, with Amazon and Clarifai performing slightly better than chance on this task. 
\begin{table}
  \small
  \caption[Summary - Acc]{Overall accuracy on designated facial analysis prediction tasks.}
  \centering
    \label{tab:over-acc}
  \begin{tabular}{l|rrrrr}
    \toprule
& Gender & Age & Name & Smile & Detection
\\
    \midrule
    Microsoft & 99.94\%&	74.09\%	&98.69\%	&79.94\%  &93.56\% \\

    Amazon & 99.75\%	&58.40\%	&87.25\%	&94.16\%  &99.25\% \\

    Clarifai & 85.97\%&	55.24\%&	95.00\%&	56.19\%  &99.31\% \\
    \bottomrule
  \end{tabular}
\end{table}

\subsubsection{Performance on Unitary Subgroups}

Tables~\ref{tab:freq_celeb_skin}--\ref{tab:freq_celeb_gender} show that with few exceptions, all APIs perform worst on the darker subgroup and female subgroup across tasks, a finding supported by previous work \cite{Joy1}. Clarifai, the only commercial API which was not previously publicly audited for the task of gender classification, demonstrated notably higher disparities across unitary groups for that task compared to the competing APIs.

\begin{table}
  \small
\caption[Summary-Unitary-Skin]{Difference in accuracy between the lighter (L) subgroup and darker (D) subgroup for each prediction task.}
 \centering
  \label{tab:freq_celeb_skin}
  \begin{tabular}{l|rrrrr}
    \toprule
    Task & Gender	&Age&	Name	&Smile & Detection
    \\
    \midrule
    Microsoft &0.13\%&	18.35\%&	1.41\%&	-0.48\%&	3.38\%  \\
    
    Amazon &0.25\%	&16.83\%&	1.03\%	&-0.75\% &	0.25\%  \\

    Clarifai & 11.69\%&	1.00\%	&7.50\%	& 0.12\% &	0.42\%\\
    \bottomrule
  \end{tabular}
\end{table}

\begin{table}
  \caption[Summary-Unitary-Gender]{Difference in accuracy between the Male (M) subgroup and female (F) subgroup for each prediction task.}
 \centering
  \small
  \label{tab:freq_celeb_gender}
  \begin{tabular}{l|rrrrr}
    \toprule
    Task & Gender	&Age&	Name	&Smile & Detection\\
    \midrule
    Microsoft &0.13\%	&9.90\%	&1.23\%	&-4.45\% & 0.62\%\\

    Amazon &0.00\%		&12.28\%		&4.75\% 	&	-9.00\% &0.50\%\\
    
    Clarifai & 	7.58\%&	10.26\%	&-1.01\%&	1.25\% &-1.63\%\\
    \bottomrule
  \end{tabular}
\end{table}

\subsubsection{Performance on Intersectional Subgroups}
Table~\ref{tab:intersect} lists intersectional subgroup performance and shows patterns found in previous work: that the most common ``least accurate subgroup'' is the darker female subgroup and the most common ``most accurate subgroup'' is the lighter male subgroup, though there are varied configurations with exceptions to this trend. 

\begin{table*}
  \small
  \caption[Summary - Intersectional]{Difference in accuracy between the best and worst performing intersectional subgroups by prediction task. The subgroups are darker females (DF), darker males (DM), lighter females (LF) and lighter males (LM). Values in bold denote equal performance. For instance, 0.25\% (DM/LM/LF - DF) signifies that the difference in accuracy between DM and DF, i.e. DM-DF, LM-DF and LF-DF are all 0.25\%.}
  \centering
    \label{tab:intersect}
  \begin{tabular}{l|rrrrr}
    \toprule
&Gender &Age &Name &Smile & Detection($AP_{50}$)\\
    \midrule
    
    Microsoft &0.25\%	\textbf{(DM/LM/LF - DF)} &	29.47\%	(LF-DF) &	3.90\%	(LF-DF)&	8.02\%	(LF-LM) & 4.25\%	(LM-DM)\\
    
    Amazon & 0.50\% (LF-DF) & 29.10\%(LM-DF) & 6.71\%	(DM-DF) &	9.75\%	(DF-LM) & 0.75\%	\textbf{(LM-DF/LF)}\\
    
    Clarifai & 19.10\%	(LM-DF)& 11.21\%	(LM-DF)&10.50\%	(LM-DF)&3.00\%	(LF-LM) & 0.50\%	\textbf{(LM/LF-DF)}\\
    \bottomrule
  \end{tabular}
\end{table*}

\section{Acknowledging and Working Through Audit Ethical Concerns}
\label{sec:method}

Using our audit conducted with CelebSET as an example, we walk through ethical concerns in current algorithmic auditing practices. We separate these concerns into design considerations and tensions. While ethical \emph{design considerations} outline additional points to be noted during the audit design in order for the audit to truthfully represent the performance of the system, ethical \emph{tensions}, represent situations where different ethical ideals come into conflict and hard decisions need to be made regarding an appropriate path forward. 

\subsection{Design Considerations}

\subsubsection{Consideration 1:  Selecting Scope of Impact}

Algorithmic audits can target a specific demographic group, prediction task, or company. This narrow scope of targets can facilitate greater impact, focusing efforts of improvement on addressing the highest risk threats. However, doing so also significantly limits the scope of the audit's impact, and allow institutions to overfit improvements to the specified tasks.

The practical reliability of the results of a benchmark also depends on the contextual and temporal relevance of the data used in evaluations to the audit use case. If it is not communicated when it is appropriate to use a benchmark, then there is no indication of when it becomes an obsolete measure of performance. This also applies for aligning the context of use of the audited system and the audit - if one demographic is under-represented in a benchmark, then it should not be used to evaluate a model’s performance on a population within that demographic. Even with intersectional considerations, there is a limit to the scope of which categories are included. 

\textbf{Illustration}
The audits conducted through CelebSET reveal that these types of external audits can only be used as an accountability mechanism within a narrow scope of influence.
For instance, Clarifai has a 19.10\% discrepancy between its best performing subgroup (lighter male) and worst performing subgroup (darker female) for the gender classification task, mirroring results from the original audits of the Gender Shades study \cite{Joy1} and demonstrating a much greater disparity compared to the difference in error rates for Microsoft and Amazon (0.25\% and 0.50\% respectively). In fact, both Amazon and Microsoft have their lowest intersectional discrepancies in gender classification, a task they have been both publicly audited for, and for which both companies released updated APIs after the disclosure of audit results. This replicates findings from the Actionable Auditing study--those that have been previously audited have smaller disparities on CelebSET, compared to those that have not been previously auditied, and thus classification bias continues to be a persistent challenge within the industry \cite{AA}. 

We found that this result holds not only for the audited company but also the task itself. In our CelebSET audit, we observed the largest difference in accuracy for Microsoft and Amazon is for the age classification task (a 29.47\% and 29.10\% discrepancy respectively between the error rates for the best performing (lighter female, and lighter male, respectively) and worst performing (darker female) subgroups). Although these companies have smaller disparities in error rates for the task of binary gender classification in response to being audited, large performance disparities are identified for other tasks. This may imply that external algorithmic audits only incentivize companies to address performance disparities on the tasks they are publicly audited for. 




Also, institutions strive to reduce performance disparities across subgroups that have been the focus of prior public audits (e.g. binary gender and skin type). Since many of the audited APIs are currently proposed for use by U.S. law enforcement~\cite{Amzlaw}, immigration~\cite{MSFTICE}, and military services~\cite{ClarifaiMaven}, the focus on performance across skin types may make sense in order to assess the risk to people of color who are over-policed and subject to additional profiling in these scenarios. 
However, there are other marginalized groups or cases to consider who may be being ignored. For instance, dysfunction in facial analysis systems locked out transgender Uber drivers from being able to access their accounts to begin work \cite{Ubertrans}. These and other issues have sparked recent work to start addressing performance disparities for this specific group \cite{scheuerman2019computers}. 

While it is important to strive for equal performance across subgroups in some tasks, audits have to be deliberate so as not to normalize tasks that are inherently harmful to certain communities. 
The gender classification task on which previously audited corporations minimized classification bias, for example, has harmful effects in both incorrect and correct classification. For example, it can  promote gender stereotypes ~\cite{gebru2019oxford}, is exclusionary  of transgender, non-binary, and gender non-conforming individuals, and threatens further harm against already marginalized individuals \cite{hamidi2018gender}.
Thus, minimizing performance disparities and investing in the improvement of that task specifically may not be the most ethical focus of impact.

\subsubsection{Consideration 2: Auditing for Procedural Fairness}


Auditing outcomes is not enough. To adequately evaluate FPTs as systems embedded in their deployed environments, we need to consider the audit as more than the final system's performance on a single benchmark. It has been well established in tax compliance that taking a procedural fairness approach to organizational audits leads to more effective evaluations. For example, studies in Australia and Malaysia sought to dissuade companies from looking for loopholes or averting tax payment--a corporate malpractice that costs states billions of dollars each year. Instead of simply looking at whatever financial documents were submitted, auditors instead evaluated the company's adherence to a tax compliance process \cite{faizal2017perception,murphy2003procedural}, by auditing the companies' internal practices and documentation development processes. The result was that institutions audited in this manner subsequently became more compliant (i.e., paid more of their taxes), and felt less intimidated by auditors. Similar findings have been discovered in employment peer review \cite{ehlen1996procedural} and computer information security \cite{enger1980computer}, revealing that inspecting adherence to a fixed and defined process for compliance standards is just as important as the result of the compliance audit itself. 

Similarly, performance disparities surfaced by FPT audits do not necessarily capture the dynamics or integrity of the engineering design processes that led to these results. In some cases, ``procedural fairness'' for machine learning (ML) systems involves interpretability methods that attempt to understand how a prediction is made. In the case of automated facial analysis tasks, an example is identifying image features that are most likely to influence the output, and ensuring that these features do not encode protected attributes such as race.
However, such a perspective constitutes a fairly constrained view as an FPT's effect on people is not limited to its prediction. The manner in which the technology is developed (e.g. were there predatory data collection practices?), the types of tests that are performed, the documentation made available, and the guardrails that are put in place are all important considerations. 
FPT evaluations such as the Face Recognition Vendor Test (FRVT) from the National Institute of Standards and Technology include some version of these qualitative considerations, such as a holistic product usability test \cite{Ref34}. But there remains a need for a comprehensive auditing framework which takes into account the end-to-end product development and deployment process.

\textbf{Illustration}
To demonstrate the biases baked into the model development and design process not captured by the CelebSET audit, we examine the diversity of the selected celebrity identities included in the APIs' model design for this task. Each of the audited APIs has a model that takes as its input an image, and outputs the name of the celebrity contained in the image. We can analyze the demographic distribution of the celebrities included in each API, in order to understand who product developers consider to be a celebrity, and how representative this selection is. 

We estimate the full list of celebrity names used in the Microsoft classifier through a publicly released dataset from the company which includes 100,000 logged celebrity identities \cite{guo2016ms}. Clarifai gives users access to the full list of 10,000 celebrities through its API, and Amazon does not make the list of included celebrity identities available in any form. We obtain each celebrity's race by matching their identity to ethnicity labels on FamousFix.com and ethniccelebs.com. Table~\ref{tab:imdb-wiki-ethnicity_x} shows the breakdown of ethnicities that are represented in Microsoft's and Clarifai's databases.  
While Clarifai includes many more Caucasian celebrity identities (74\% of  celebrity labels) than any other group, Microsoft, with 37\% Caucasian, 19\% Asian and 21\% Black celebrity names included appears to have a more inclusive design. 

Had we focused solely on the performance of these APIs on CelebSET, we would have missed this label selection bias and remained with an incomplete understanding of the design flaws that influence the APIs' performance. While comprehensive auditing frameworks examining model development and deployment processes are yet to be developed, documentation proposals such as datasheets \cite{gebru2018datasheets}, model cards \cite{ModCards} and factsheets \cite{hind2018increasing} can encourage designers to carefully think about these processes if they are required elements of such an audit.

\begin{table*}
  \centering
  \small
  \caption[Breakdown of celebrity identities in commercial APIs by ethnicity.]{Breakdown of celebrity identities in commercial APIs by ethnicity.}
  \label{tab:imdb-wiki-ethnicity_x}
  \begin{tabular}{l|rrrrrrrr}
    \toprule
& Asian & White & Hispanic & Black  & Middle Eastern & Indian & Other/Mixed & Total \\
    \midrule
    Microsoft API & 7,838 & 15,536 & 10 & 8,816 & 995 & 1,316 & 7,167 & 41,678 \\
    & 18.8\% & 37.3\% & 0.02\% & 21.1\% & 2.4\% & 3.2\% & 17.2\% & 100.0\%\\
    \midrule
    Clarifai API & 172 & 4,861 & 0 & 534 & 31 & 125 & 800 & 6,523 \\
    & 2.6\% & 74.5\% & 0.0\% & 8.2\% & 0.5\% & 1.9\% & 12.3\% & 100.0\%\\
    \bottomrule
  \end{tabular}
\end{table*}

\subsection{Ethical Tensions}

\subsubsection{Tension 1: Privacy and Representation}

\begin{table*} \small 
  \centering
  \label{5}
  \caption[Breakdown of IMDB-WIKI Data Examples by Ethnicity]{Breakdown of IMDB-WIKI data examples by ethnicity.}
  \label{tab:imdb-wiki-ethnicity}
  \begin{tabular}{l|rrrrrrrr}
    \toprule
& Asian & White & Hispanic & Black & Middle Eastern & Indian & Other/Mixed & Total \\
    \midrule
    IMDB-WIKI-Eth & 7,557 & 338,896 & 351 & 29,613 & 1,160 & 3,299 & 33,468 & 414,344 \\
    & 1.8\% & 81.8\% & 0.1\% & 7.2\% & 0.3\% & 0.8\% & 8.1\% & 100.0\%\\
    \bottomrule
  \end{tabular}
\end{table*}

\begin{table}
  \small
  \centering
  \label{6}
  \caption[Breakdown of IMDB-WIKI Data Examples by Gender]{Breakdown of IMDB-WIKI data examples by gender.}
    \label{tab:imdb-wiki-gender}
  \begin{tabular}{lrrrr}
    \toprule
& Male & Female & Unknown & Total \\
    \midrule
    IMDB-WIKI & 230,912 & 179,900 & 3,532 & 414,344 \\
& 55.7\% & 43.4\% & 0.85\% & 100\%\\
    \bottomrule
  \end{tabular}
\end{table}





While audit benchmark datasets should reflect the populations who will be impacted by the audited technology, collecting a sufficiently large and diverse dataset can present privacy risks for the individuals represented in the dataset. 
Depending on data storage and dissemination policies, sensitive and biometric information may be made accessible beyond the intended auditing purpose. These risks can be further compounded by potential consent violations during the data collection process. For example, IBM's Diversity in Faces dataset was sourced from Creative Common licensed images uploaded to Flickr \cite{2019arXiv190110436M}. While these images are open for public internet use, the Flickr users who uploaded the photos, and the individuals in the photos, did not consent to being included in a facial recognition dataset \cite{IBMfail}.




Privacy and consent violations in the dataset curation process often disproportionately affect members of marginalized communities. Benchmark dataset curation frequently involves supplementing or highlighting data from a specific population that is underrepresented in previous datasets. Efforts to increase representation of this group can lead to tokenism and exploitation, compromise privacy, and perpetuate marginalization through population monitoring and targeted violence \cite{hamidi2018gender,hoffmann2019fairness,Chinasurveil}.
And the method through which companies pursue better representation can be ethically questionable. For instance, a startup signed a deal with the Zimbabwe government to harvest the faces of millions of citizens through unprecedented access to their CCTV cameras, smart financial systems, airport, railway, and bus station security, and a national facial database~\cite{Chinasurveil2}. Without seeking the active consent of impacted individuals and working towards mutual benefit, such an act can be exploitative and tokenizing of the humans contributing to the improvement of the system for which their data is used. 

\textbf{Illustration}
CelebSET was sourced from IMDB-WIKI \cite{imdbwiki}, a dataset with significant demographic bias that can be seen in the distribution of meta-data labels in Table \ref{tab:imdb-wiki-ethnicity} and Table~\ref{tab:imdb-wiki-gender}. Certain groups are not only highly underrepresented, but there are fewer images per person from these groups. This skew is a result of media and social biases that make certain subgroups less photographed than others, and thus less likely to exist in CelebSET's source data. This lack of representation becomes even more stark when considering intersectional identities such as Black women. 
 

Consequently, when aiming to design a ``demographically balanced'' benchmark, i.e., one with equal representation from the designated demographic subgroups, it is more challenging to sufficiently represent certain groups relative to others. Thus, by actively seeking to include members of underrepresented groups, the privacy risk is disproportionately increased for that group. For instance, in this case, there is twice the likelihood that an image from the ``Black'' subgroup of the reference dataset will be included in CelebSET than an image from the ``White'' subgroup.

 
In addition to the ethical challenges emerging from the skewed distribution of the data source, we also encounter challenges pertaining to obtaining consent. The CelebSET dataset is sourced from a database of public figures. While these individuals have, from a legal perspective, opted in to having their likeness used freely in the public domain, we acknowledge that they have not consented to inclusion in an FPT benchmarking dataset specifically. While an opt-in informed consent process would be ideal for subjects included in the benchmark, the individuals in CelebSET are effectively unreachable and thus cannot be contacted to give informed opt-in consent. Even the less ideal opt-out model is challenging to implement as many subjects may never become aware that their face is in the dataset, thus rendering the option to opt-out meaningless. The consistency of the benchmark can also be compromised if the dataset changes over time through the removal of individuals. 

\subsubsection{Tension 2: Intersectionality and Group-Based Fairness}

\begin{figure}
\centering
\includegraphics[scale=0.55]{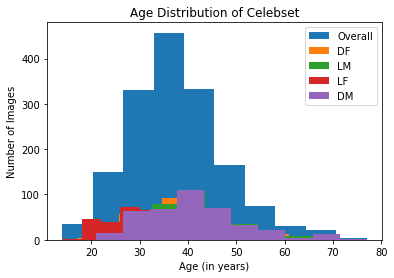}
\caption{Histogram of the age distribution in CelebSET. The median age is 37, the mode is 36 and the mean is 37.56. The youngest subject is 14 and the oldest is 77. Blue signifies overall age distribution. Purple, orange, green and red show the age distribution for darker males, lighter males, darker females and lighter females respectively.}
\centering
\label{age}
\end{figure}







The concept of intersectionality, coined by legal scholar Crenshaw \cite{crenshaw1989demarginalizing}, is a framework for understanding how interlocking systems of power and oppression give rise to qualitatively different experiences for individuals holding multiply marginalized identities \cite{Crenshaw2017}. 


Crenshaw writes of moving beyond stereotypical assignments and recognizing decision outcomes on a more individualized basis -- observing an individual to be at the intersection of numerous unique combinations of identities, possessing several dimensions of privilege and oppression. However, in order for group fairness to be evaluated, an individual's experience must be reduced to a categorical assignment, even while performing disaggregated analysis to account for multiple categories. 
Although inspired by intersectionality, this type of multi-axis disaggregated analysis fails to capture how systems of power and oppression give rise to qualitatively different experiences for individuals holding multiply marginalized identities \cite{hoffmann2019fairness}.  



\textbf{Illustration}
While developing our CelebSET benchmark, we paid careful attention to balance it with respect to a crawled ``ethnicity'' label and binary gender. Although such a design is derived from the naturally occurring labels of the crawled and referenced datasets, the selected groupings have inherent limitations. Unlike the Pilot Parliaments Benchmark from the Gender Shades study \cite{Joy1} where the intersectional groups are defined with respect to skin type, ``ethnicity'' is an attribute that is highly correlated but not deterministically linked to racial categories, which are themselves nebulous social constructs, encompassing individuals with a wide range of phenotypic features \cite{Benthall:2019:RCM:3287560.3287575}. Similarly, binary gender labels are compatible with the format of commercial product outputs, but exclusionary of those not presenting in the stereotypical representations of each selected gender identity \cite{Keyes:2018:MMT:3290265.3274357}. 


We can see in Figure~\ref{age} that although CelebSET is balanced with respect to gender and coarse ethnicity labels, it is highly unbalanced with respect to age.  At times, the exclusion of a particular group is unavoidable - for instance, our dataset lacks children because of the legal restrictions around the online exposure of a child. The average age of darker females in our benchmark is 37, with lighter males at 39, lighter females at 32, and darker males at 41. Since there are many more younger lighter females than older darker males, it is unclear whether the disparities between those groups is more correlated with age rather than race or gender. It is thus possible to optimize performance across gender and ethnicity categories in our benchmark, while continuing to perform poorly with respect to age (i.e. improve accuracy for only older darker males). This is an observation of the fairness gerrymandering effect \cite{Kearns2018PreventingFG} - where optimizing for fairness on one axis can compromise fairness in another.


\subsubsection{Tension 3: Transparency and Overexposure}





To limit misinterpretations of evaluation results on specific benchmarks, it is important to clearly communicate the limit of each benchmark and its appropriate context of use.
Sharing details of the dataset development process with auditors and targets helps clarify the limit of the audit's scope, and the context in which results should be interpreted and appropriately acted upon. 
Similarly, publicly disclosing named audit targets can incite pressure to make the audit itself more impactful \cite{AA}. 
However, all this may come at great cost - such communications can also lead to targets overfitting to optimize product performance on the audit.
Audits of this nature have also made institutions wary -- in September 2019, IBM, an audit target in the Gender Shades study, removed its facial recognition capabilities from its publicly distributed API \cite{IBM0919}. Similarly, Kairos began putting its services behind an expensive paywall following its inclusion in the Actionable Auditing study \cite{Kairosprice19}. Such practices, although rightfully stopping developers from using a product revealed to be flawed, also compromise the product's auditability -- making it more expensive and challenging for auditors to evaluate, even though it may still be in active use by enterprise customers.

\textbf{Illustration}
In order to communicate the biases and limitations of CelebSET, we can create a datasheet \cite{gebru2018datasheets} which helps clarify the context in which the benchmark should be used. This datasheet can specify, for instance, which demographic groups are covered in the audit, and what types of product applications the benchmark is best suited for. We can additionally note the small scope and limited demographic groupings of this particular audit.  
However, IMDB-Wiki, from which CelebSET and its variants are derived, is a publicly available online dataset \cite{imdbwiki}. This means the entire process of benchmark development is easily accessible to anyone including an audit target, who may decide to include the images in their training set. It is thus inevitable that the dataset will become obsolete as products overfit on the data. 

\section{Reconsidering the Role of Algorithmic Audits}

Our work shows that the algorithmic audit itself is a testing ground for the ethical concerns it is meant to evaluate. The audits need to be done with careful attention to the traps their targets fall into, and auditors must strive to live up to the ethical ideals they expect from their targets.  

Auditors, thus, should approach these evaluations with a certain level of humility, acknowledging the limitations of their own evaluations and contextualizing each benchmark result as one component of a larger and more qualitative audit framework, which should begin by questioning the ethical use case of the product itself. The humble goal of the algorithmic audit is thus to expose blind spots rather than validate performance. Given its very own limitations and ethical concerns, FPT audits on benchmarks like CelebSET are a necessary but insufficient condition. They are inspections that can be used to stall or halt deployment, but do not have the weight of meaning to, by themselves, be used to justify FPT deployment or act as a condition for a moratorium. This means CelebSET as a benchmark should not be considered as a reward to game or a goal to strive for, but a very low bar not to be caught tripping over.




\section{Conclusion}
\label{sec:conclusion}
While designing CelebSET, an audit process for products employing facial processing technology (FPT), we were able to identify several ethical concerns with the developing norms of the algorithmic auditing of such products. 
These concerns in audit outcomes and processes often intersect with those of the audited product itself, as an unethical audit process can lead to a false sense of progress on the alignment of facial processing technology with the principles we have put forth. 
Both the audit process and the audited FPT, for instance, need to have careful privacy considerations, and avoid exploiting marginalized groups in the blind pursuit of increasing representation. 
If we take seriously the ethical expectations we have for the audited product, then we must also apply that same standard to the data and processes defining our evaluation.

\bibliographystyle{ACM-Reference-Format}
\balance
\bibliography{FRT_bib}
\end{document}